\documentclass[12pt,preprint]{aastex}

\shorttitle{Detection of the Nitrosylium Ion}
\shortauthors{Cernicharo et al.}

\usepackage{graphicx}
\usepackage{amsmath}
\usepackage{subfigure}
\usepackage{natbib}
\usepackage{multirow}
\usepackage{txfonts}
\begin{document}
\title{Tentative Detection of the Nitrosylium Ion in Space
\thanks{This work was based on observations carried out with the 
IRAM 30-meter telescope. IRAM is supported by INSU/CNRS (France), 
MPG (Germany) and IGN (Spain)}}

\shorttitle{Discovery of Nitrosylium in Space}
\shortauthors{J. Cernicharo et al.}

\author
{
J. Cernicharo\altaffilmark{1,2}, 
S. Bailleux\altaffilmark{3},
E. Alekseev\altaffilmark{4}
A. Fuente\altaffilmark{5},
E. Roueff\altaffilmark{6},
M. Gerin\altaffilmark{7}, 
B. Tercero\altaffilmark{1,2}, 
S.P. Trevi\~no-Morales\altaffilmark{8},
N. Marcelino\altaffilmark{9,10},
R. Bachiller\altaffilmark{5},
B. Lefloch\altaffilmark{11}
}

\altaffiltext{1}{Departamento de Astrof\'isica, Centro de Astrobiolog\'ia, CSIC-INTA, Ctra. de Torrej\'on a Ajalvir km 4,28850 Madrid, Spain}
\altaffiltext{2}{ICMM. CSIC. Group of Molecular Astrophysics. C/Sor Juana In\'es de la Cruz N3. 28049 Cantoblanco. Madrid, Spain}
\altaffiltext{3}{Laboratoire de Physique des Lasers, Atomes et Mol\'ecules, UMR 8523 CNRS, Universit\'e Lille 1, 59655 Villeneuve d'Ascq Cedex, France.}
\altaffiltext{4}{Institute of Radio Astronomy, National Academy of Sciences of Ukraine, Krasnoznamennaya ul. 4, Kharkov 61002, Ukraine}
\altaffiltext{5}{Observatorio Astron\'omico Nacional, Apdo. 112, 28803, Alcal\'a de Henares, Spain}
\altaffiltext{6}{LERMA, Observatoire de Paris, CNRS UMR 8112.  Place J. Janssen, 92190 Meudon, France.}
\altaffiltext{7}{LERMA, Observatoire de Paris, CNRS UMR8112 and Ecole Normale Superieure, 61 Avenue de l'Observatoire, F-75014 Paris, France}
\altaffiltext{8}{Instituto de Radio Astronom\'ia Milim\'etrica (IRAM), Avenida Divina Pastora 7, Local 20, 18012 Granada, Spain}
\altaffiltext{9}{NRAO, 520 Edgemont Road, Charlottesville, VA22902, USA}
\altaffiltext{10}{Italian ALMA Regional Centre, Instituto di Radioastronomia, Via P. Gobeti 101, 40129 Bologna, Italy}
\altaffiltext{11}{UJF-Grenoble/CNRS-INSU, Institut de Plan\'etologie et d'Astrophysique de Grenoble (IPAG) UMR 5274, 38041 Grenoble, France}

   \date{Received ...; accepted ...}

\begin{abstract}
We report the tentative detection 
in space of the nitrosylium ion, NO$^+$. The observations
were performed towards the cold dense core Barnard 1-b. 
The identification of the NO$^+$ $J$=2--1 line is supported by new laboratory measurements of NO$^+$ rotational 
lines up to the $J$=8--7 transition (953207.189\,MHz), which leads to an improved set of molecular constants:
$B_0 = 59597.1379(62)$\,MHz, $D_0 = 169.428(65)$\,kHz, and $eQq_0(\textrm{N}) = -6.72(15)$\,MHz.
The profile of the feature assigned to NO$^+$ exhibits two velocity components at 6.5 
and 7.5 km s$^{-1}$, with column densities of $1.5 \times 10^{12}$ and $6.5\times10^{11}$ cm$^{-2}$, respectively.
New observations of NO and HNO, also reported here, allow to estimate the following abundance ratios:
$X$(NO)/$X$(NO$^+$)$\simeq511$, and $X$(HNO)/$X$(NO$^+$)$\simeq1$. This latter value provides important 
constraints on the formation and destruction processes of HNO.
The chemistry of NO$^+$ and other related nitrogen-bearing species is investigated by the means of a time-dependent gas phase model 
which includes an updated chemical network according to recent experimental studies. 
The predicted abundance for NO$^+$ and NO is found to be consistent with the observations.
However, that of HNO relative to NO is too high.
No satisfactory chemical paths have been found to explain the observed low abundance of HNO. 
HSCN and HNCS are also reported here with an abundance ratio of $\simeq1$.
Finally, we have searched for NNO, NO$_2$, HNNO$^+$, and NNOH$^+$, but only
upper limits have been obtained for their column density, except for the latter for which 
we report a tentative 3-$\sigma$ detection.
\end{abstract}


\vspace{0.5cm}
To appear in the Astrophysical Journal October 20, 2014
\vspace{0.5cm}

\section{Introduction}

Nitrogen is one of the major contributors to the chemical richness of the interstellar medium (ISM), and it
is able to form a large variety of molecules in the gas phase,
including HCN, HNC, CN, and NH$_3$.
The apolar  N$_2$ molecule, probably the most abundant N-bearing molecular in the interstellar gas, 
is nearly unobservable directly by using conventional spectroscopic techniques, so its protonated ion
diazenylium, N$_2$H$^+$, is usually used as a tracer of N$_2$. 
Diazenylium has been observed towards many different astronomical sources 
\citep{Turner1974,Green1974,Fuente1993,Daniel2006,Daniel2007,Daniel2008,Daniel2013}, 
while ammonium, NH$_4^+$, has been only detected towards the cold dark core B1-b \citep{Cernicharo2013}.
The study of all these N-bearing species is of paramount importance to understand interstellar chemistry, 
in particular for studying the formation of N-bearing organic species, which play a crucial role in the chemistry relevant to life.

From a chemical point of view, the cold dark core B1-b is characterized by a peculiar molecular profuseness. 
Several molecules like HCNO \citep{Marcelino2009}, CH$_3$O \citep{Cernicharo2012}, and NH$_3$D$^+$ \citep{Cernicharo2013}, have been observed in this object for the first time.  Also peculiar is its high degree of deuterium fractionation as shown by the presence
of multiply deuterated molecules, such as ND$_3$ \citep{Lis2002,Lis2010} or D$_2$CS \citep{Marcelino2005}. 
Several complex organic molecules, typical of hot cores or corinos, have also been reported towards B1-b
\citep{Oberg2010, Nuria2007, Cernicharo2012}. 
Because of its rich chemistry and  its great interest for the comprehension of the star formation process, B1-b was targeted
within the Large Program \lq{}Astrochemical Surveys at IRAM\rq{} (ASAI, Lefloch \& Bachiller, 2014, in preparation). 

Interstellar molecules containing nitrogen and oxygen simultaneously have been poorly studied until now.
Nitric oxide has been detected toward a number of star-forming clouds \citep[][and references therein]{Liszt1978,Gerin1992}. 
\citet{Halfen2001} searched for NO$^+$ and NO$_2$ towards SgrB2(N) without success.

In this work we report on the discovery
for the first time in space of the
nitrosylium ion, NO$^+$, together with observations of several lines of other molecules
containing the NO group. New laboratory experiments have been performed in order to
improve the rotational constants of nitrosylium. The derived abundances
are discussed in the context of time dependent models of the gas phase Nitrogen chemistry.  
Although at low frequencies B1 exhibits a large number of lines, its low kinetic temperature
produces a significant reduction of the line intensities at high frequency. Hence,
its spectral density above 200\,GHz is very low \citep{Cernicharo2013}.
The detection of the nitrosylium ion
is based on just one line but it is the only feature, together with the transition 
1$_{11}$-0$_{00}$ of $o$-SD$_2$, appearing over 1.8 GHz of 
bandwidth and just at the predicted frequency of the J=2-1 line of NO$^+$.

\begin{figure}
\begin{center}
\includegraphics[angle=0,scale=.61]{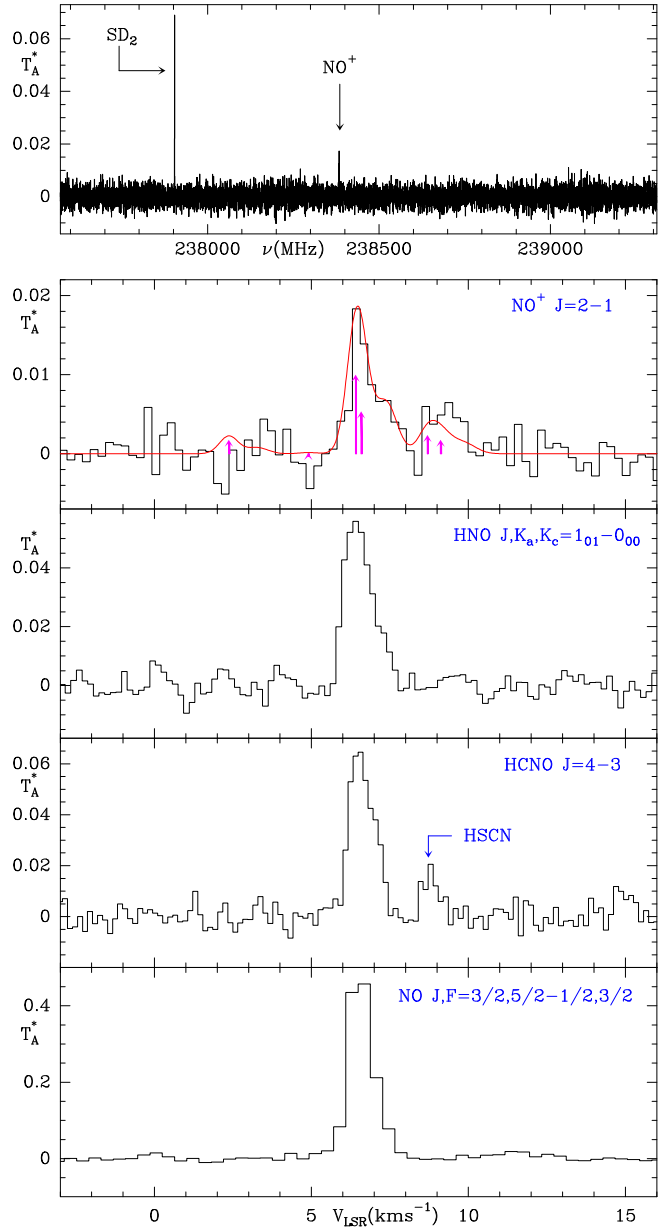}
\caption{Observed spectrum of the $J$=2-1 line of NO$^+$ ({\it two top panels}) towards the cold core B1-b. 
The first panel shows the complete observed spectrum with 1.8 GHz of frequency coverage around 238.4\,GHz. 
Only two lines are
unambigously detected in the data belonging to SD$_2$ and NO$^+$. The second top panel shows a zoom to the NO$^+$ J=2-1 transition
with its six hyperfine
components indicated by vertical arrows for the 6.5 km\,s$^{-1}$ velocity component (see text). 
Relatives intensities of the different components are proportional to the length of the arrows (from left to right,
0.5, 0.033, 2.8, 1.5, 0.5, and 0.667).
The continuous line (in red) corresponds to the predicted intensity of NO$^+$
from a model with two velocity components (see text).
Selected lines from HNO, NO and HCNO are also shown in the other panels.
The abcissa scale is V$_{LSR}$ in km\,s$^{-1}$ (except for the top panel for which the scale is in MHz) 
and that of the ordinate is antenna temperature in Kelvin
(a color figure is available in the online version).
} 
\label{Fig1}
\vspace{-0.5cm}
\end{center}
\end{figure}

\section{Observations}
The observations presented in this paper are part of a complete spectral line survey at
3, 2 and 1\,mm of the cold core B1-b ($\alpha_{J2000}$=03$^{\rm h}$ 33$^{\rm m}$ 
20.8$^{\rm s}$, $\delta_{J2000}$=31$^\circ$ $07' 34''$) within the above mentioned IRAM large program ASAI. 
All observations were performed at the IRAM 30-m radiotelescope at Pico Veleta (Spain)
using the Eight MIxer Receivers (EMIR) and the fast Fourier Transform Spectrometers (FTS) with a spectral 
resolution of 198 kHz,  during several runs in 2012 and 2013. The observations were carried out in wobbler switching mode, 
with a throw of $\pm$120{\arcsec}, providing excellent baselines. These data were
complemented with a high spectral resolution (50 kHz) frequency sweep at 3\,mm obtained
between 2003-2007, and January and March 2012 \citep[][and in prep.]{Marcelino2005,Marcelino2007,Marcelino2009,Cernicharo2012}.
These 3\,mm high spectral resolution observations were performed in Frequency Switching 
mode with a frequency throw of 7.14 MHz, which removes standing waves between the secondary 
and the receivers.

Weather conditions were mostly average winter to moderate summer for all observations, 
with 1--10 mm of precipitable water vapor. System temperatures were in the range 80--300 K.
Each frequency setup was observed for $\sim2$ hrs, with pointing checks
in between on strong and nearby quasars. Pointing errors were always within 3$''$. 
The 30m beam sizes are between $30''-21''$ at 3\,mm, $20''-17''$ at 2\,mm, and
$12''-9''$ at 1\,mm. The spectra were calibrated in antenna temperature corrected for atmospheric 
attenuation using the ATM package \citep{Cernicharo1985,Pardo2001}. Figure 1 shows a selection of the 
observed spectra and Table~\ref{Table:1} provides the line parameters for all the observed lines of NO$^+$, 
NO, HNO and HCNO.
For each observed line the error quoted for T$_A^*$ corresponds to the noise sensitivity per 
spectral channel. The adopted main beam efficiencies are 0.8, 0.73, 0.58, and 0.54 for frequencies 81-115, 150-165, 238, and 
250\,GHz, respectively. Main beam antenna temperature for the observed lines
can be obtained from the antenna temperature by dividing by the corresponding main beam efficiencies. 

The first ASAI observing runs at 1mm started in January 2013 and revealed a 3-$\sigma$ feature just at the 
expected frequency
of the $J$=2-1 line of NO$^+$. In order to improve its signal-to-noise ratio, we performed additional observations 
in September, November and
December 2013. The final averaged spectrum, with a noise rms level of 2 mK per 200\,KHz channel (total observing time of
43 hours on-source), is shown
in Figure 1 together with selected observed lines of the related species HNO, NO, and HCNO.
The top panel of this figure shows the complete observed spectrum at 238.4 GHz and covering 1.8 GHz of bandwidth. It shows only
two lines, one from SD$_2$ and the other one right at the predicted frequency of the J=2-1 transition
of NO$^+$.
During these observations lines from unrelated species such as HSCN were also detected (see Figure~\ref{Fig1}). 
All data have been processed using the GILDAS software package 
(http://www.iram.fr/IRAMFR/GILDAS/).

The $J$=2-1 transition of NO$^+$ has several hyperfine components (see Table~\ref{Table:1}, ~\ref{Table:2},
Figure~\ref{Fig1} and Section 3). The strongest pair 
($F$=2-1 \& 3-2), at an averaged frequency of 238383.17 MHz,  is separated
by only 140 kHz, so it is not spectrally resolved by our observations.
Another hyperfine line should be present at a velocity of -4.1 km\,s$^{-1}$ (relative to the velocity
of the source, 
see Figure~\ref{Fig1}). Its relative intensity with respect the strongest one is 1/8.6, so it is well below our detection limit. 
Finally, another pair of hyperfine components ($F$=2-2 \& 1-0),  separated by 340 kHz, should be present at +2.4 km\,s$^{-1}$ 
with an intensity relative to the strongest component of 1/3.7. As Figure~\ref{Fig1} shows, 
a spectral feature is well detected at the expected position of this pair of lines.

The density of spectral lines in B1-b is rather low at the frequency of the NO$^+$ $J$=2-1 transition
\citep[see][]{Cernicharo2013} but, nevertheless, we searched for alternative line candidates in the MADEX code \citep{Cernicharo2012b}, 
which contains spectroscopic information for more than 5013 spectral entries, and the CDMS and JPL catalogs \citep{Muller2005,
Pickett1998}, 
and we found no plausible molecular lines, other than that of NO$^+$, able 
to explain the observed features. This feature is the only one appearing over the 1.8 GHz spectrum shown in
Figure~\ref{Fig1} (together with a transition of SD$_2$). The difference between the observed
frequency and that predicted from laboratory data is less than 20 KHz. This
could make us confident with the assignment of the observed feature to the nitrosylium ion. 
Unfortunately, the second strongest group of hyperfine
components is detected just above 3$\sigma$ level. Hence, we consider our assignment of the observed feature 
to NO$^+$ as a tentative detection.

\begin{table*}
\caption[]{Observed Line Parameters}
\vspace{-0.4cm}
\begin{center}
\resizebox{1.0\textwidth}{!}{
\begin{tabular}{|c | c| c| c| c| c| c| c| c| c|}
\hline
Molecule & Transition                       & $\nu_{rest}$     & E$_{upp}$ & A$_{ij}$   & S$_{ij}$   & $\int$T$_A^*$dv        & V$_{LSR}$         & $\Delta$v    &T$_A^*$\\
         &                                  &   (MHz)          &  (K)      & (s$^{-1}$) &            & (K\,km\,s$^{-1}$)& (km\,s$^{-1}$)    & (km\,s$^{-1}$)& (K)  \\
\hline
NO$^+$&     2 2 - 1  1           &  238383.111( 35) & 17.2 & 6.130 10$^{-6}$ & 1.500 &           &        &         &       \\
      &     2 3 - 1  2           &  238383.255( 36) & 17.2 & 8.174 10$^{-6}$ & 2.800 &  0.011( 2)& 6.43(8)& 0.58(12)& 0.017(2)\\
      &                          &                  &      &           &       &  0.008( 2)& 7.23(9)& 0.95(30)& 0.008(2)\\
      &     2 2 - 1  2           &  238381.099( 83) & 17.2 & 2.043 10$^{-6}$ & 0.500 &           &        &         &         \\
      &     2 1 - 1  0           &  238381.435( 73) & 17.2 & 4.541 10$^{-6}$ & 0.667 &  0.007( 2)& 6.92(9)& 1.11(24)& 0.006(2)\\
      &     2 1 - 1  1           &  238386.464(130) & 17.2 & 3.406 10$^{-6}$ &       &           &        &         & $<$0.006\\
\hline   
HNO   &     1$_{01}$ - 0$_{00}$  &   81477.490(100) &  3.9 & 2.226 10$^{-6}$ & 1.000 &  0.069( 2)& 6.48(2)& 1.10( 4)& 0.059(3)\\                                  
HNO   &     2$_{02}$ - 1$_{01}$  &  162937.949(100) & 11.7 & 3.561 10$^{-5}$ & 2.000 &  0.070( 5)& 6.58(3)& 0.88( 8)& 0.074(6)\\
HNO   &     2$_{12}$ - 1$_{11}$  &  159802.294(100) & 36.1 & 2.520 10$^{-5}$ & 1.500 & $<$0.013( 7)& 6.68(9)& 0.71(15)& 0.017(9)\\
\hline     
NO   &  3/2 -1 5/2 - 1/2  1 3/2 &  150176.4566( 7) &  7.2 & 3.310 10$^{-7}$ & 2.000 &  0.494( 7)& 6.56(1)& 0.92( 2)& 0.504(8)\\
NO   &  3/2 -1 3/2 - 1/2  1 1/2 &  150198.7573( 7) &  7.2 & 1.839 10$^{-7}$ & 0.740 &  0.182( 7)& 6.55(2)& 0.87( 4)& 0.195(8)\\
NO   &  3/2 -1 3/2 - 1/2  1 3/2 &  150218.7421( 7) &  7.2 & 1.471 10$^{-7}$ & 0.592 &  0.174( 8)& 6.56(2)& 0.99( 6)& 0.165(8)\\
NO   &  3/2 -1 1/2 - 1/2  1 1/2 &  150225.6504( 7) &  7.2 & 2.943 10$^{-7}$ & 0.592 &  0.151( 6)& 6.57(2)& 0.84( 4)& 0.169(7)\\
NO   &  3/2  1 3/2 - 1/2 -1 3/2 &  150439.0941( 7) &  7.2 & 1.480 10$^{-7}$ & 0.593 &  0.168( 6)& 6.52(2)& 1.02( 5)& 0.155(7)\\
NO   &  3/2  1 5/2 - 1/2 -1 3/2 &  150546.4622( 7) &  7.2 & 3.334 10$^{-7}$ & 2.000 &  0.529( 5)& 6.54(1)& 0.95( 1)& 0.525(6)\\
NO   &  3/2  1 1/2 - 1/2 -1 1/2 &  150580.5531( 7) &  7.2 & 2.964 10$^{-7}$ & 0.592 &  0.162( 7)& 6.52(2)& 0.94( 5)& 0.162(7)\\
NO   &  3/2  1 3/2 - 1/2 -1 1/2 &  150644.3489( 7) &  7.2 & 1.853 10$^{-7}$ & 0.739 &  0.189( 7)& 6.48(2)& 0.94( 4)& 0.189(7)\\
NO   &  5/2  1 7/2 - 3/2 -1 5/2 &  250436.8416(11) & 19.2 & 1.841 10$^{-6}$ & 3.200 &  0.229( 3)& 6.47(1)& 0.67( 1)& 0.323(3)\\
     &                                      &                  &      &           &       &  0.100( 3)& 6.99(1)& 0.75( 4)& 0.126(3)\\
NO   &  5/2  1 5/2 - 3/2 -1 3/2 &  250440.6532(11) & 19.2 & 1.547 10$^{-6}$ & 2.020 &  0.126( 3)& 6.43(2)& 0.61( 3)& 0.193(3)\\
     &                                      &                  &      &           &       &  0.081( 4)& 6.90(2)& 0.68( 6)& 0.111(3)\\
NO   &  5/2  1 3/2 - 3/2 -1 1/2 &  250448.5255(11) & 19.2 & 1.381 10$^{-6}$ & 1.200 &  0.129( 2)& 6.58(2)& 0.87( 2)& 0.138(3)\\
NO   &  5/2  1 3/2 - 3/2 -1 3/2 &  250475.4186(11) & 19.2 & 4.420 10$^{-7}$ & 0.384 &  0.038(12)& 6.63(6)& 0.84( 5)& 0.043(3)\\
NO   &  5/2  1 5/2 - 3/2 -1 5/2 &  250482.9387(11) & 19.2 & 2.947 10$^{-7}$ & 0.384 &  0.034(12)& 6.49(9)& 0.68(12)& 0.047(3)\\
NO   &  5/2 -1 5/2 - 3/2  1 5/2 &  250708.2416(11) & 19.3 & 2.958 10$^{-7}$ & 0.384 &  0.036(12)& 6.58(3)& 0.90( 8)& 0.037(3)\\
NO   &  5/2 -1 3/2 - 3/2  1 3/2 &  250753.1342(11) & 19.3 & 4.437 10$^{-7}$ & 0.384 &  0.022(18)& 6.41(9)& 0.50(14)& 0.040(3)\\
     &                                      &                  &      &           &       &  0.010(19)& 6.93(9)& 0.58(13)& 0.017(3)\\
NO   &  5/2 -1 7/2 - 3/2  1 5/2 &  250796.4225(11) & 19.3 & 1.849 10$^{-6}$ & 3.200 &  0.259( 3)& 6.52(2)& 0.73( 1)& 0.334(3)\\    
     &                                      &                  &      &           &       &  0.061( 2)& 7.08(9)& 0.75( 6)& 0.077(3)\\
NO   &  5/2 -1 5/2 - 3/2  1 3/2 &  250815.6097(11) & 19.3 & 1.554 10$^{-6}$ & 2.020 &  0.214( 3)& 6.58(2)& 0.92( 3)& 0.222(3)\\
NO   &  5/2 -1 3/2 - 3/2  1 1/2 &  250816.9300(11) & 19.3 & 1.387 10$^{-6}$ & 1.200 &  0.121( 3)& 6.58(2)& 0.82( 2)& 0.140(3)\\
\hline
NNO  & 4 - 3                    &  100491.7196(11) & 12.1 & 1.358 10$^{-7}$ & 4.000 &  0.003( 1)& 6.56(15)& 0.61(15)&0.005(2)\\
NNO  & 6 - 5                    &  150735.0453(16) & 25.3 & 4.759 10$^{-7}$ & 6.000 &           &         &         &$<$0.021\\
\hline
HCNO  &     4 - 3                &   91751.3120(41) & 11.0 & 3.838 10$^{-5}$ & 4.000 &  0.046( 7)& 6.47(5)& 0.65( 8)& 0.067(3)\\
      &                          &                  &      &           &       &  0.016( 7)& 7.06(8)& 0.50(13)& 0.031(3)\\
HCNO  &     5 - 4                &  114688.3827(47) & 16.5 & 7.667 10$^{-5}$ & 5.000 &  0.033( 3)& 6.56(3)& 0.64( 6)& 0.049(8)\\
HCNO  &     6 - 5                &  137624.9341(50) & 23.1 & 1.345 10$^{-4}$ & 6.000 &  0.023( 3)& 6.74(7)& 1.00(16)& 0.022(3)\\
HCNO  &     7 - 6                &  160560.8717(50) & 30.8 & 2.160 10$^{-4}$ & 7.000 &  0.020( 7)& 7.0(2) & 1.09(35)& 0.018(8)\\
\hline
NNOH$^+$ & 4$_{04}$ - 3$_{03}$      &   89541.4026(63) & 10.7 & 1.073 10$^{-5}$ & 4.000 &  0.008( 2)& 6.71(9)& 0.91(21)& 0.009(3)\\
NNOH$^+$ & 5$_{05}$ - 4$_{04}$      &  111925.3586(71) & 16.1 & 2.144 10$^{-5}$ & 5.000 &           &        &         & $<$0.014\\
NNOH$^+$ & 6$_{06}$ - 5$_{05}$      &  134308.3849(75) & 22.6 & 3.761 10$^{-5}$ & 6.000 &           &        &         & $<$0.018\\
NNOH$^+$ & 7$_{07}$ - 6$_{06}$      &  156690.2956(75) & 30.1 & 6.039 10$^{-5}$ & 7.000 &           &        &         & $<$0.024\\
\hline
HNNO$^+$ & 4 - 3  $F$=5-4          &   94369.817( 46) & 11.3 & 4.508 10$^{-5}$ & 4.890 &           &        &         & $<$0.021\\
HNNO$^+$ & 4 - 3  $F$=4-3          &   94369.838( 46) & 11.3 & 4.226 10$^{-5}$ & 3.750 &           &        &         & $<$0.021\\
HNNO$^+$ & 4 - 3  $F$=3-2          &   94369.897( 46) & 11.3 & 4.140 10$^{-5}$ & 2.860 &           &        &         & $<$0.021\\
\hline
\end{tabular}
}
\end{center}
\vspace{-0.2cm}
{Predicted rest frequencies, Einstein coefficients and line strengths are from the MADEX code \citep{Cernicharo2012b}.
Quantum numbers are $J$, and $F$ for NO$^+$; $J$, $K_a$, and $K_c$ for HNO and NNOH$^+$; $J$ for HCNO; $J$ , $p$ (parity),
and $F$ for NO ($\Omega=1/2$); $J$, $K_a$=0, $K_c$=$J$, and $F$ for HNNO$^+$. The number in parentheses correspond to the 
uncertainty referred to the last digits of the 
predicted or observed values. Upper limits for the reported line parameters correspond to 3\,$\sigma$ values.}
\label{Table:1}
\end{table*}

\section{Laboratory Spectroscopy}

While the nitrosyl radical has a $^2\Pi$ electronic ground state, that of nitrosylium is $^1\Sigma^+$.
The dipole moment of NO$^+$ has not been measured in the laboratory, but it has been calculated with quantum 
chemistry computations to be $\simeq$0.36 D
\citep{Jungen1970,Billingsley1973,Chambaud1990,Polak2004}.
High resolution spectroscopic data of NO$^+$ are rather scarce. Two pure rotational transitions with resolved nitrogen quadrupolar
hyperfine structure transitions have been measured in the millimeter-wave band {\citep{bowman82a}}.
In the infrared  region, a total of eight lines of the fundamental band have been observed using 
diode laser spectroscopy {\citep{Ho91a,Hilpert94a}}.
In addition, about fourty ro-vibrational transitions of the fundamental and first hot bands have been recorded with
the MIPAS interferometer onboard the EnviSat satellite {\citep{Lopez2006}}.
The $A$ $^1\Pi$ - $X$ $^1\Sigma^+$ emission bands of NO$^+$ have been studied by \cite{Alberti1975}.

In view of the limited number of spectroscopic data available in the  millimeter-wave range, we decided 
to extend the laboratory measurements
of rotational transitions of NO$^+$ up to frequencies just below 1 THz.
It is worth mentioning that the high atmospheric opacity prevents detecting interstellar NO$^+$ at the lowest rotational
transition $J$=1-0 (119.19\,GHz) from ground-based observations due to a molecular oxygen line falling nearby.

Details of the submillimeter-wave spectrometer employed in this study have been described previously 
by {\cite{Ozeki2011}}. Briefly speaking, it consists of a 2 m free space, double-jacketed Pyrex cell allowing liquid-nitrogen flow through the
outer jacket, a liquid He-cooled InSb hot electron bolometer, and a submillimeter-wave source.
The submillimeter-wave radiation between 580 and 955 GHz was provided by three phase-locked backward wave oscillators 
(Istok Company).

The nitrosylium ion was produced in a magnetically extended ($\sim$200G) negative glow dc discharge in NO employed as precursor.
The discharge current maintained in the plasma was adjusted to 8 mA.
We found that the plasma did not need to be cooled using liquid nitrogen flow, and that
the use of a buffer gas such as argon or helium was not necessary.

Vibrational excitation of the ion was high enough to detect rotational lines in vibrational states up to $v = 2$.
We extended the measurements to the $^{15}$NO$^+$ isotopologue, which was also observed in vibrationally excited states.
Data obtained for these two isotopic species were sufficient to perform an isotopically invariant fit using the generalised
Dunham expansion of the energy levels. The corresponding results will be published elsewhere, and we focus here on the conclusions 
relative to the main species observed in its ground vibrational state.

\begin{figure}
\begin{center}
\includegraphics[angle=0,scale=.95]{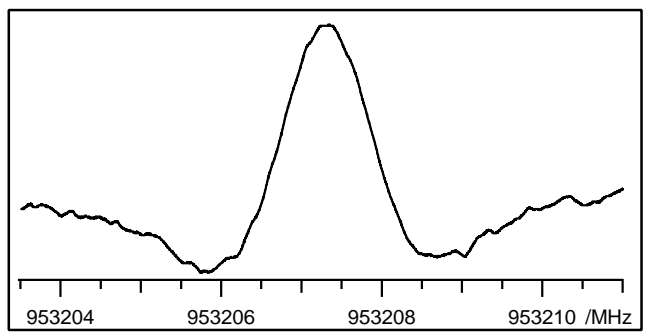}
\caption{
Observed $J$=8-7 rotational transition of $^{14}$N$^{16}$O$^+$ in $\nu$=0 using NO as precursor. The line
was recorded by integrating 16 scans with 0.5 Hz repetition rate and 10\,ms time constant of the
lock-in amplifier.
} 
\label{Fig2}
\vspace{-0.5cm}
\end{center}
\end{figure}

NO$^+$ is a close-shell molecule and the electric quadrupole moment of Nitrogen ($I_{^{14}{\mathrm{N}}} = 1$) 
splits rotational energy levels with $J > 0$
in three hyperfine sublevels denoted by $F = \{J, J \pm 1$\}. For $J=0$ the only possible nuclear spin sublevel 
corresponds to $F = 1$.
Note that due to the Doppler-limited linewidth of the high-$J$, newly measured transitions prevented the observation of the 
hyperfine-structure components.
Prediction and least-squares analysis of observed millimeter-wave transitions have been carried out using Pickett's CALPGM package 
{\citep{Pickett91}}.
In total, four new transitions have been observed in the range 595-954 GHz, corresponding to $4 \le J'' \le 7$.
Figure~\ref{Fig2} shows the spectral line due to the $J = 8 \leftarrow 7$ transition near 953 GHz.
Table~\ref{Table:2} lists the frequencies of all transitions measured in the millimeter-wave range.
The derived molecular constants are (MHz) $B_{0}$=59597.1379(62), $D_{0}$=0.169428(65), and
$eQq$=-6.72(15). The least-squares fitting procedure gave a standard deviation of 80 kHz.
As expected, this new determination of the NO$^+$ molecular parameters significantly improves
the previous determination by \cite{bowman82a} ($B_{0}$=59597.132(16), 
$D_{0}$=0.171(1), $eQq$=-6.76(10))

\begin{table}
\caption[]{Laboratory transition frequencies of $^{14}$NO$^+$.}
\begin{center}
\vspace{-0.4cm}
\begin{tabular}{|c |c |c |r|}
\hline
  $J' \rightarrow J''$ & $F' \rightarrow F''$ & $\nu_{\mathrm{obs}}$ (MHz) & $\Delta\nu$$^{\rm a}$ (MHz) \\
\hline
                       $2 \rightarrow 1$   &  $1 \rightarrow 0$   & \multirow{2}{*}{238381.200$^{\rm b}$}  &  \multirow{2}{*}{$-0.106$}  \\
                                          &  $2 \rightarrow 2$   &                            &            \\
                                          &  $2 \rightarrow 1$   & \multirow{2}{*}{238383.200$^{\rm b}$}  &  \multirow{2}{*}{$-0.023$}  \\
                                          &  $3 \rightarrow 2$   &                            &            \\
                                          &  $1 \rightarrow 1$   &      238386.430$^{\rm b}$  &  $-$0.058  \\
                       $3 \rightarrow 2$   &  $4 \rightarrow 3$   &      357564.320$^{\rm b}$  &  $-$0.173  \\
                                          &  $3 \rightarrow 2$   &      357564.320$^{\rm b}$  &  $-$0.173  \\
                                          &  $2 \rightarrow 1$   &      357564.320$^{\rm b}$  &  $-$0.173  \\
                       $5 \rightarrow 4$   &                     &      595886.721$^{\rm c}$  &  $+$0.066  \\
                       $6 \rightarrow 5$   &                     &      715019.297$^{\rm c}$  &  $+$0.034  \\
                       $7 \rightarrow 6$   &                     &      834127.480$^{\rm c}$  &  $-$0.004  \\
                       $8 \rightarrow 7$   &                     &      953207.189$^{\rm c}$  &  $-$0.038  \\
\hline
\end{tabular}
\end{center}
{$^{\rm a}$ $\Delta\nu = \nu_{obs}-\nu_{calc}$.
$^{\rm b}$ {\cite{bowman82a}}.
$^{\rm c}$ This work.}
\label{Table:2}
\end{table}

\section{Results and Discussion}
The astronomical data were analyzed using the MADEX code \citep{Cernicharo2012b}.
Taking into account the low value of the dipole moment of NO$^+$, 0.36 D 
\citep{Polak2004}, we could expect
the population of its rotational levels to be very close to thermal equilibrium in B1-b. 
At a distance of $\sim$300 pc, the Barnard 1 (B1) dark cloud, which is part of the large scale complex of molecular clouds in 
Perseus \citep{Bachiller1984,Bachiller1990}, comprises several dense cores, at different evolutionary stages of star formation.
While B1-a and B1-c are known to host class 0 sources, with developed outflows \citep{Hatchell2005}, the B1-b core
appears to be more quiescent.
B1-b consists of two very dense cores, B1-bN and B1-bS, separated by 20'' 
\citep{Huang2013} which are considered to be 
protostellar sources even younger than Class 0 protostars, possibly First Hydrostatic Cores (FHSC) according to \cite{Pezzuto2012}.
A more evolved third source, named B1-b-Spitzer because it was identified in Spitzer infrared data, exhibits deep absorption features 
from ices \citep{Jorgensen2006}. The three sources are deeply embedded in a protostellar envelope of large column 
density, N(H$_2$)=10$^{23}$ cm$^{-2}$, large volume density, n(H$_2$)$\simeq$10$^5$ cm$^{-3}$ \citep{Daniel2013}, and low kinetic temperature, 
T$_K \simeq$  12-15 K,  which seems essentially unaffected by the internal sources 
\citep{Marcelino2005,Marcelino2009,Lis2010}. Hence, most low-$J$ lines of diatomic and triatomic molecules could
exhibit rotational temperatures close to the kinetic temperature of the gas.


Assuming a rotational temperature of 12 K, we derive a column density for NO$^+$ of (1.5$\pm$0.5)\,$\times$10$^{12}$ cm$^{-2}$ for
the velocity component at 6.5 km\,s$^{-1}$ which is the one exhibiting a large number of
complex molecules \citep{Oberg2010,Cernicharo2012,Cernicharo2013}.  A second velocity component, at 7.2-7.5 km\,s$^{-1}$, is
detected in spectral lines of several molecular  species  \citep[see, e.g.,][]{Marcelino2005,Oberg2010,Huang2013}. For this
component we obtain a column density of 
(6.5$\pm$3)$\times$10$^{11}$ cm$^{-2}$. The combined column density of NO$^+$ over the two velocity components 
is $\simeq$(2.2$\pm$0.8)$\times$10$^{12}$ cm$^{-2}$.

The observations of HNO and NO have a lower spectral resolution than that of 
NO$^+$ (except for the HNO 1$_{01}$-0$_{00}$
line at 3mm), so the two velocity components are blended in the corresponding line profiles. However, the
measured linewidths are larger than those of the 3\,mm window for which we used a spectral
resolution of 49\,kHz. From the parameters listed in Table~\ref{Table:1}, we derive 
a column density for HNO and NO of (2$\pm$0.5)$\times$10$^{12}$ and
(1.1$\pm$0.2)$\times$10$^{15}$ cm$^{-2}$, respectively. Indeed these values have to be considered as corresponding to 
the average of the two velocity components. Hence, the abundance ratios between the three NO-bearing species
NO, NO$^+$ and HNO are: $X$(NO)/$X$(NO$^+$)$\simeq$511, $X$(NO)/$X$(HNO)$\simeq$550,
and $X$(NO$^+$)/$X$(HNO)$\simeq$1.

Another interesting molecule containing the NO group is fulminic acid, HCNO, which was previously observed by
\cite{Marcelino2009}. The line parameters for the HCNO lines observed in our line survey are reported in 
Table~\ref{Table:1}. The HCNO column density derived from these new data, 2$\times$10$^{11}$ cm$^{-2}$,
is in very good agreement with the results of \cite{Marcelino2009} who discussed the chemistry of this species.

We have also searched in our line
surveys at 3 and 2\,mm for the $J$=4-3 and $J$=6-5 lines of NNO. A 2.5-$\sigma$ feature has been observed just at the frequency
of the $J$=4-3 transition (see Table~\ref{Table:1}). The 3-$\sigma$ upper limit to the column density of NNO is
4$\times$10$^{12}$ cm$^{-2}$. A search for the 1$_{11}$-0$_{00}$ strongest hyperfine component of NO$_2$ at 253338.12\,MHz
(E$_{upper}$=12.2 K) provides only a 3-$\sigma$ upper limit to its column density of 
5$\times$10$^{12}$ cm$^{-2}$. We have searched for the two related cations NNOH$^+$ and
HNNO$^+$. For the latter isomer of N$_2$OH$^+$, which 
is less stable by 6\,kcal/mol \citep{Javahery1990}, we derive
an upper limit (3-$\sigma$) to its column density of 1.5$\times$10$^{11}$ cm$^{-2}$. For the more stable isomer, NNOH$^+$, we have 
observed a feature for its 4$_{04}$-3$_{03}$ transition (see Table~\ref{Table:1})
at 3-$\sigma$ level. The corresponding column density is $\simeq$10$^{11}$ cm$^{-2}$. However,
more sensitive observations are needed to confirm the detection of this ion for the first time in space.

Figure 1 shows the 8$_{08}$-7$_{07}$ line of HSCN. This molecule was detected by \cite{Halfen2009} towards SgrB2(N) and towards
TMC1 by \cite{Adande2010}. From the observed intensity and linewidth of this transition
(T$_{MB}$=0.021 K, $\Delta$v=0.8 km\,s$^{-1}$),
we derive a column density of (2$\pm$0.5)$\times$10$^{11}$\,cm$^{-2}$. 
HNCS, first detected in space by \citep{Frerking1979}, was also detected in our survey. 
From the observed line parameters for its 7$_{07}$-6$_{06}$ transition 
(T$_{MB}$=0.014\,K, $\Delta$v=0.8 km\,s$^{-1}$), we derive N(HNCS)=(2.5$\pm$0.9)$\times$10$^{11}$\,cm$^{-2}$. 
Hence, the abundance ratio $X$(HNCS)/$X$(HSCN) in B1-b is found to be $\sim$1, similar to the value estimated in  
TMC1 by \cite{Adande2010}, and in SgrB2(N) by \cite{Halfen2009}.
Such low abundance ratios between two isomers separated by more than 3000 K in energy suggest
that they are formed from a common ion precursor, probably HSCNH$^+$ \citep{Halfen2009,Adande2010}.




\begin{figure}
\includegraphics[angle=0,scale=0.6]{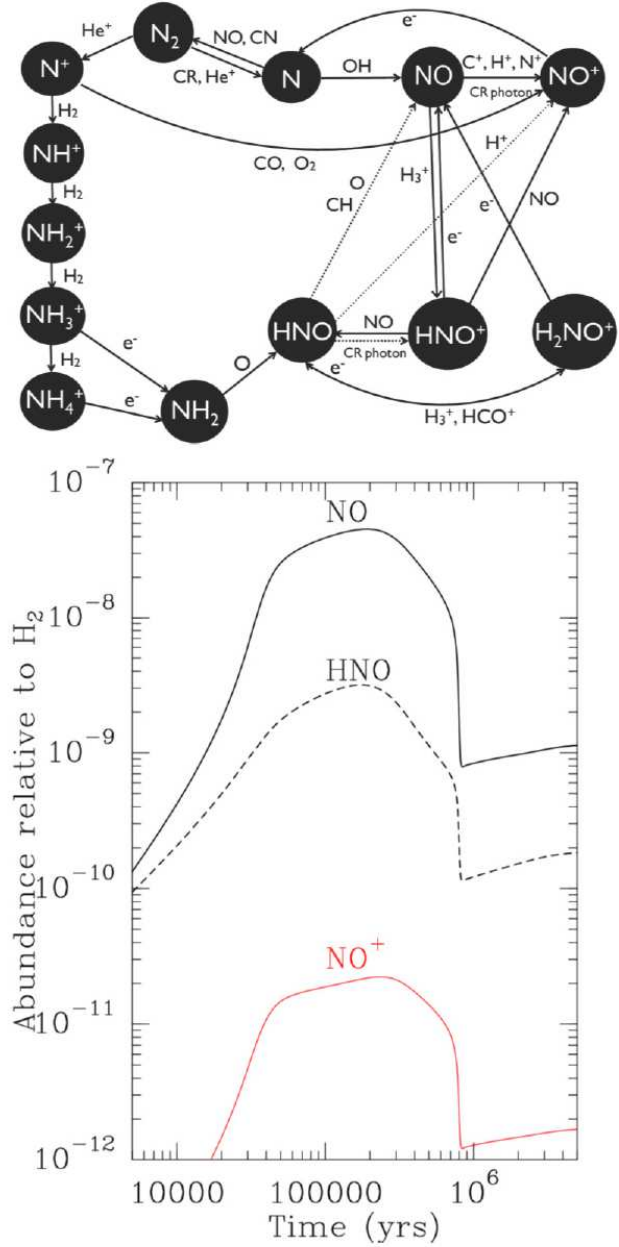} 
\caption{The top panel shows the chemical network
used in our models for NO, NO$^+$ and HNO. The bottom panels show
the abundances of NO$^+$, NO and HNO relative to H$_2$ versus time (see
text). }
\label{Fig3}
\end{figure}


The formation and destruction mechanisms of NO$^+$ in low and
high-density interstellar clouds have been discussed by \citet{Herbst1973},\citet{Pickles1977}, 
and by \citet{Singh1980}.
Briefly summarized (see Fig.~\ref{Fig3}), possible pathways in cold dense interstellar clouds include 
charge transfer reactions between NO and various ions (C$^+$, H$^+$, CH$_3^+$, ...), exothermic ion molecule 
reactions between N$^+$ and oxygen containing species such as CO, O$_2$, NO, ..., and the reaction of HNO with H$^+$, He$^+$. 
Since NO$^+$ does not react with H$_2$ \citep{Huntress1977}, the dominating (almost unique) mechanism responsible 
for its destruction is the dissociative recombination reaction 
which produces back oxygen and nitrogen atoms. The reaction NO$^+$ + e$^-$ $\rightarrow$ N + O has been well studied both 
in laboratory and through theoretical ab-initio studies with a very satisfactory agreement \citep{Schneider2000,Sheehan2004}.
 
We have considered a time dependent gas phase model including an updated nitrogen chemistry, following the 
recent experimental studies of \cite{Daranlot2012,Daranlot2013} reported in \cite{Wakelam2013}. The chemical 
network contains 117 chemical species and 1117 chemical reactions. Figure~\ref{Fig3} displays the 
time evolution of NO$^+$, NO, and HNO for typical dark cloud conditions (n(H$_2$) = 10$^5$ cm$^{-3}$, 
T = 10\,K). 
The fractional elemental 
abundances adopted are 0.1, 7$\times$10$^{-6}$, 2$\times$10$^{-5}$, 10$^{-5}$, 
and 1.5$\times$10$^{-8}$ for He, C, O, N, and M (metals), respectively, which have been used in previous chemical studies for 
deuterated molecules in the B1 cloud \citep{Marcelino2005,Lis2010,Cernicharo2013}. 
The abundance of NO$^+$ is almost linearly dependent on the cosmic ionization rate. The elemental 
abundance of sulfur has been somewhat varied to better match the observations. We show in Figure~\ref{Fig3} the results 
obtained with an elemental abundance of sulfur of 10$^{-5}$, as the sulfur molecules have 
been found relatively abundant in B1 \citep{Marcelino2007}, and a cosmic ionization rate of 2$\times$10$^{-17}$ s$^{-1}$ per H$_2$ molecule. 

The predicted abundance for NO$^+$
at t=2$\times$10$^5$ yr is consistent with observations, $X$(NO$^+$)=
2.2$\times$10$^{-11}$ versus
the observed value of $\simeq$2$\times$10$^{-11}$. For the same time our gas phase model predicts $X$(NO)=
4.6$\times$10$^{-8}$,
and $X$(NO)/$X$(NO$^+$)$\simeq$2100. These values are 4 times larger than the observed ones. Moreover,
our model also fails to reproduce the observed HNO abundance, which is similar to that of NO$^+$, 
by a factor $\simeq$1-2 10$^2$.
Neutral-neutral reactions play a significant role in the nitrogen chemistry as first emphasized by \cite{Forets1990}. The NO radical is principally formed 
through N + OH, and HNO results from the O + NH$_2$ reaction. The corresponding rates, recently updated by \cite{Wakelam2013}, have been introduced
in the model. We did not include two nitrogen containing compounds such as N$_2$O, and HN$_2$O$^+$ which could be formed from NO and 
HNO. Including the corresponding additional reactions is desirable but unfortunately, very little information is available for these species, 
either experimentally or theoretically.
The detection of NO$^+$ and the comparison of its abundance with related species, such as NO and HNO, provides
important constraints on the gas phase chemical models. More developed and thorough models, including
chemical processes on the surfaces of grains, and additional gas phase reactions of NO and HNO with other neutral species,
are certainly needed to better understand the chemistry of N- and O-bearing molecules.

\acknowledgements
We would like to thank Spanish MINECO for funding support under grants 
CSD2009-00038,  AYA2009-07304, AYA2012-32032, and FIS2012-32096.
The National French Program "Physique et Chimie du Milieu Interstellaire" (PCMI) 
is acknowledged for its funding support.  E. A. thanks the University of Lille and CNRS for
funding support. The French National Research Agency (ANR-13-BLAN "IMOLABS") is also acknowledged.
The National Radio Astronomy Observatory is a facility of the National
Science Foundation operated under cooperative agreement by Associated
Universities, Inc.
The online database KIDA (Wakelam et al. 2012, http://kida.obs.u-bordeaux1.fr) has been consulted for 
checking some reaction rate coefficients used in the chemical model.

\end{document}